\documentclass[fleqn,usenatbib]{mnras}

\usepackage{newtxtext}

\usepackage[T1]{fontenc}

\DeclareRobustCommand{\VAN}[3]{#2}
\let\VANthebibliography\thebibliography
\def\thebibliography{\DeclareRobustCommand{\VAN}[3]{##3}\VANthebibliography}

\usepackage{graphicx}	
\usepackage{amsmath}	

\newcommand{\msun}{$M_{\odot}$}	
\newcommand{\td}{\mathrm{d}}	

\title[GW parameter estimation with neural networks]{Fast Bayesian gravitational wave parameter estimation using convolutional neural networks}

\author[M. Andrés-Carcasona et al.]{
M. Andrés-Carcasona,$^{1}$\thanks{E-mail: mandres@ifae.es}
M. Martínez,$^{1,2}$
Ll. M. Mir$^{1}$
\\
$^{1}$Institut de Física d'Altes Energies (IFAE), The Barcelona Institute of Science and Technology, Campus UAB, E-08193 Bellaterra (Barcelona), Spain\\
$^{2}$Catalan Institution for Research and Advanced Studies (ICREA), E-08010 Barcelona, Spain
}

\date{Accepted XXX. Received YYY; in original form ZZZ}

\pubyear{2015}

\begin{document}
\label{firstpage}
\pagerange{\pageref{firstpage}--\pageref{lastpage}}
\maketitle

\begin{abstract}
The determination of the physical parameters of gravitational wave events is a fundamental pillar in the analysis of the signals observed by the current ground-based interferometers. Typically, this is done using Bayesian inference approaches which, albeit very accurate, are very computationally expensive. We propose a convolutional neural network approach to perform this task. The convolutional neural network is trained using simulated signals injected in a Gaussian noise. We verify the correctness of the neural network's output distribution and compare its estimates with the posterior distributions obtained from traditional Bayesian inference methods for some real events. The results demonstrate the ability of the convolutional neural network to produce posterior distributions that are compatible with the traditional methods. Moreover, it achieves a remarkable inference speed, lowering by orders of magnitude the times of Bayesian inference methods, enabling real-time analysis of gravitational wave signals. Despite the observed reduced accuracy in the parameters, the neural network provides valuable initial indications of key parameters of the event such as the sky location,  facilitating a multi-messenger approach. 
\end{abstract}




\section{Introduction}

Gravitational waves (GW) are ripples in the fabric of spacetime produced by moving massive objects such as binary black holes or rotating neutron stars. These waves were not directly detected until the recent discovery by the LIGO and Virgo collaborations in 2015 \citep{FirstGWDet}. Since then, a total of 90 events have been detected in the three publicly released catalogs \citep{Abbott2019GWTC-1:Runs,Abbott2020GWTC-2:Run,Abbott2021GWTC-3:Run}, corresponding to the O1-O3 observing runs of Advanced LIGO \citep{AdvLIGO} and Advanced Virgo \citep{AdvVIRGO}. The fourth observing run started in Spring 2023, already detecting tenths of new events. The majority of them are binary black hole mergers, but binary neutron star mergers and black hole-neutron star mergers have also been detected. The discovery of GWs has opened up a new window to study the universe, as they can be used, for example, to infer the population of merging compact objects \citep{POP}, test the theory of General Relativity \citep{TGR} or understand the composition of neutron stars \citep{CW1,CW2,CW3}.

In order to extract useful information from the detected GW signals, it is necessary to accurately perform an estimation of the parameters of the source leading to the observed signal. Typically, this is done using a Bayesian inference framework, in which a probability distribution is constructed over the parameter space based on the observed data and prior knowledge \citep{Veitch_LALinference,Singer:2015_RapidBayesian,Abbott2019GWTC-1:Runs,Bilby,Bilby2,Smith:2019rapidBayesian}. Markov Chain Monte Carlo (MCMC) or Nested Sampling (NS) algorithms are commonly used to wander the parameter space and sample the posterior probability distribution \citep{Gilks_MCMC,NestedSampling,Veitch_LALinference,Bilby2}. These methods, albeit powerful and precise, can be computationally intensive and do not scale appropriately with the increasing number of events. Due to the large volume of expected detections of future runs, in particular in next-generation experiments, new tools or implementations are being studied \citep{Berry:2014,Pankow:2015,Singer:2015_RapidBayesian,cuoco2020enhancing,Bhardwaj2023,Alvey:2023naa,Crisostomi_2023}.

Recently, there has been a growing interest in using artificial intelligence and machine learning techniques to address this task. Among those, convolutional neural networks (CNNs), normalizing flows and autoencoders have been the most common approaches followed \citep{George:2017_DL,Gabbard:2017,Fan:2018,Gabbard:2019,Green:2020,Green:2020GW150914,Krastev:2020}. CNNs have been applied to detect events \citep{George:2017_DL,Menendez-Vazquez2021SearchesPeriod,Morras:2021,Andres-Carcasona:2022}, to perform the parameter estimation of the signals \citep{Chua:2019,Gabbard:2019,Green:2020} or to classify glitches \citep{GWspy,George:2018}, among other applications (see \cite{cuoco2020enhancing} for a comprehensive review). 

The best performing machine learning algorithms applied to the parameter estimation problem are the ones described by \cite{Gabbard:2019,DaxDingo,Green:2020,Green:2020GW150914}. \cite{Gabbard:2019} use a variational autoencoder to predict the posterior probability distribution typically produced by a Bayesian inference approach. Their results are almost identical to those obtained using a Bayesian inference technique but to generate $8,000$ samples only takes $0.1$s, an improvement of several orders of magnitude. A different approach is followed by \cite{DaxDingo}, where they estimate the Bayesian posterior using a neural network and then modify the distribution using importance sampling. This method takes between one and ten hours (depending on the waveform used) to perform the parameter estimation running on a GPU and multiple cores, which is a great improvement over the several hours to days that typically take the MCMC samplers to wander the parameter space. Finally, \cite{Green:2020,Green:2020GW150914} use normalizing flows to produce accurate posterior distributions, comparable to those produced by traditional methods.

In this paper, we present a new and fast method for GW parameter estimation using a CNN. This article is organized as follows. The training data set and preprocessing is explained in Sec. \ref{sec:dataset}. In Sec.~\ref{sec:architecture} the CNN architecture is presented. The training procedure is explained in Sec.~\ref{sec:training}. Finally, Sec.~\ref{sec:results} shows the results obtained for the test set and for three real events, comparing it to the results obtained with the traditional MCMC approach.

\section{Dataset and preprocessing}
\label{sec:dataset}

To generate the training data for the CNN we first take realizations of Gaussian noise that follow the power spectral density (PSD) of the strain measured for the O3b run in the three interferometers \citep{Abbott2021GWTC-3:Run}. Then, we simulate and inject on this noise GW signals using the IMRPhenomv2 waveform \citep{IMRPhenomPv2_1,IMRPhenomPv2_2} from the PyCBC library \citep{nitz2020gwastro,Usman2016TheCoalescence}. The parameters of the injected signals are those described in Tab.~\ref{tab:Parameters}. The delay between the different interferometers and its antenna patterns are also taken into account.

\begin{table}
	\centering
	
	\begin{tabular}{lccc} 
		\hline \hline
		\textbf{Parameter} & \textbf{Symbol} & \textbf{Distribution} & \textbf{Units}\\
		\hline \hline
		Mass 1 & $m_1$ & Uniform$(20, 60)$ & [\msun]\\
		Mass 2& $m_2$ & Uniform$(20, 60)$ & [\msun]\\
		Distance & $d$ & Uniform$(100, 3000)$ & [Mpc] \\
        Right ascension & $\alpha$ & Uniform$(0,2\pi)$ & [rad] \\
        Cosine of declination & $\cos(\delta)$ & Uniform$(-1,1)$ & - \\
        Polarization angle & $\psi$ & Uniform$(0,\pi)$ & [rad] \\
        Inclination & $\theta_{JN}$ & Uniform$(0,\pi/2)$ & [rad] \\
        Time of coalescence & $t_c$ & Uniform$(0.5, 0.9)$ & [s] \\
        Spin magnitude 1 & $a_1$ & Uniform$(-1, 1)$ & - \\
        Spin magnitude 2 & $a_2$ & Uniform$(-1, 1)$ & - \\
		\hline \hline
	\end{tabular}
 	\caption{Parameter space of the training set. We additionally assume that $m_1\geq m_2$.}
          \label{tab:Parameters}

\end{table}

The strain is then whitened and normalized by dividing it by its standard deviation. We choose a sampling frequency of $4096$ Hz and include the information of LIGO Hanford, LIGO Livingstone and Virgo at the same time. The frequency is restricted to the $(30~\mathrm{Hz},1024~\mathrm{Hz})$ range, as the signals fit well within these values. The signal is then cropped to $1$s  containing the merger between the time $0.5$ s and $0.9$ s (the exact coalescing time is taken as random over a uniform distribution). Therefore, the input matrix for the CNN has a shape of $(4096,3)$, each column corresponding to the data of each interferometer. 

The signal-to-noise ratio (SNR) is defined as 
\begin{equation}
    \mathrm{SNR} = \sqrt{4\int_{f_{\min}}^{f_{\max}}\td f \frac{|\tilde{h}(f)|^2}{S_n(f)}}~,
\end{equation}
where $\tilde{h}(f)$ is the strain in frequency domain and $S_n(f)$ is the PSD of the strain noise. The network SNR is then computed as
\begin{equation}
    \mathrm{SNR}_{\mathrm{net}} = \sqrt{\sum_{i= 1}^3\mathrm{SNR}_i^2}~,
\end{equation}
where $i$ runs over the three interferometers. We restrict ourselves to signals that satisfy $\mathrm{SNR}_\mathrm{net}\geq 10$.

A total of $601,600$ signals are used for the training stage, $38,400$ for the validation and $100$ are kept for testing the performance afterwards. 

An improved efficiency has been observed when fitting some derived quantities rather than the original ones presented in Tab.~\ref{tab:Parameters}. For instance, instead of directly estimating the individual masses of the black holes, we utilize the chirp mass, denoted by $\mathcal{M}_c$ and which is calculated using the expression

\begin{equation}
\mathcal{M}_c = \frac{(m_1m_2)^{3/5}}{(m_1+m_2)^{1/5}}~,
\end{equation}
alongside the mass ratio, denoted by $q$ and computed as
\begin{equation}
    q = \frac{m_1}{m_2}~,
\end{equation}
where $m_1\geq m_2$. 

Similarly, instead of predicting the individual spins, we use the effective spin, defined as

\begin{equation}
    \chi_{\mathrm{eff}} = \frac{a_1m_1+a_2m_2}{m_1+m_2}~.
\end{equation}

Furthermore, we make the decision not to fit the polarization and inclination angles, as they are not among the key parameters required for real-time analyses. As a result, the CNN will estimate the set of parameters $\pmb{\theta}=\left\{\mathcal{M}_c,q,d,t_c,\chi_{\mathrm{eff}},\alpha,\delta \right\}$.




\section{CNN architecture}
\label{sec:architecture}

\begin{figure*}
	\includegraphics[width=\textwidth]{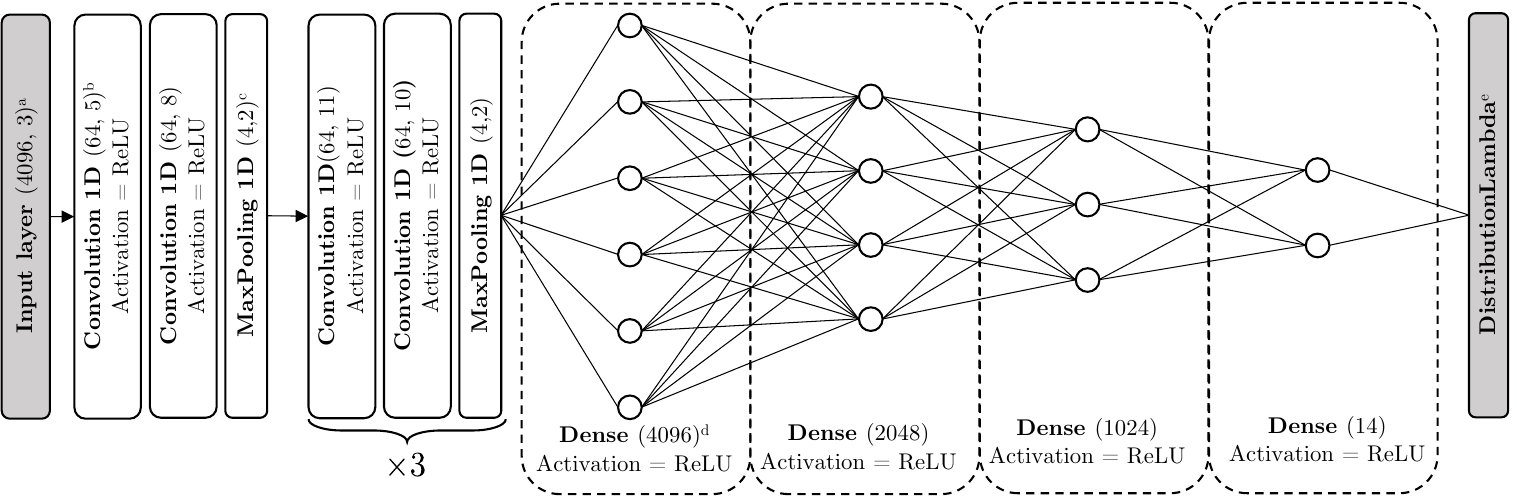}
    \caption{Architecture of the convolutional neural network used in this work. $^a$ Size of the input layer. $^b$ Filters and size of the kernel of the 1D convolution layer. $^c$ Size of the pool size of the 1D maximum pooling layer. $^d$ Number of neurons of the fully connected dense layer. $^e$ Layer that includes the 7 Kumaraswamy distribution (one per each estimated parameter). Before this layer, the exponential of the numbers outputted by the last dense fully connected layer are taken.}
    \label{fig:CNNarch}
\end{figure*}

The CNN architecture that has been used in this work is displayed in Fig.~\ref{fig:CNNarch}. The first set of layers are one-dimensional convolutions and one-dimensional max pooling ones, followed by a set of fully connected dense layers. Between each of the dense layers, a dropout $20\%$ is applied to prevent overfitting during the training stage. 

To be able to capture the uncertainty in the final estimation of the parameters we choose to model the full distribution as the posterior in Bayesian statistics would, instead of producing point estimates. Therefore, the last dense layer that contains $14$ neurons will be connected to a TensorFlow Probability layer with a Kumaraswamy distribution for each parameter. 

The Kumaraswamy distribution has a probability density function defined by \citep{Kumaraswamy}
\begin{equation}
    f(x;\alpha,\beta) = \alpha\beta x^{\alpha-1}(1-x^\alpha)^{\beta-1}~,
\end{equation}
for $x\in[0,1]$, which is a flexible enough distribution. It is very similar to the beta distribution, but its simple expression of the probability density function allows to have an easy evaluation of the quantile function, which takes the form of
\begin{equation}
    F^{-1}(\xi; \alpha,\beta) = (1-(1-\xi)^{1/\beta})^{1/\alpha}~.
\end{equation}

Sampling from this distribution is then trivial by applying the inverse transform sampling method. If we draw $\xi\sim \mathrm{Uniform}(0,1)$ and then compute $F^{-1}(\xi; \alpha,\beta)$ for each realization, we will obtain the desired samples of this distribution. This procedure is very fast in contrast to sampling from 
a distribution for which no analytical expression of the quantile function is available or which is computationally expensive to evaluate. 

Since our variables are not in the range $[0,1]$ , the estimated parameters need to be transformed as
\begin{equation}
    x =\frac{\theta - \theta_{\min}}{\theta_{\max}-\theta_{\min}}~,
\end{equation}
where $\theta\in \pmb{\theta}$.

The numbers estimated for the different variables are different in orders of magnitude and, therefore, we  estimate the $\log \alpha$ and $\log \beta$ instead. This motivates taking the exponential of the 14 numbers generated by the last dense layer before evaluating the probability distribution.

The loss function that will be used is the negative log-likelihood (i.e. $L = -\log \mathcal{L}(\alpha,\beta | X)$). For the Kumaraswamy distribution and $n$ observations, the log-likelihood equals
\begin{equation}
\begin{aligned}
        \log \mathcal{L} &=  n\log \alpha +n\log\beta+(\alpha-1)\sum_{i= 1}^{n}\log(x_i)\\ &+(\beta-1)\sum_{i=1}^n\log(1-x_i^\alpha)~.
\end{aligned}
\end{equation}

Choosing this loss function is equivalent to maximizing the log-likelihood, meaning that the neural network is actually learning to find the parameters of the distribution that best fit the data.

\section{Training}
\label{sec:training}

To build and train the CNN we use Keras with TensorFlow's backend and its implementation in GPUs \citep{abadi2016tensorflow}. For managing the probabilistic layers we use distribution layers \citep{TensorflowProbability} from TensorFlow Probability. 

We feed the data in batches of 32 signals as was done in \cite{Menendez-Vazquez2021SearchesPeriod,Andres-Carcasona:2022}. The learning rate is set to $10^{-5}$. The metrics tracked during this stage are mainly the loss and the validation loss. The former is computed with the training set and represents the function being minimized, while the latter is computed using the validation set. This allows to observe the appearance of overfitting or underfitting during the training procedure. The training lasts for 8 epochs and the one yielding a minimum validation loss is going to be the neural network used for inference.

\begin{figure}
	\includegraphics[width=\columnwidth]{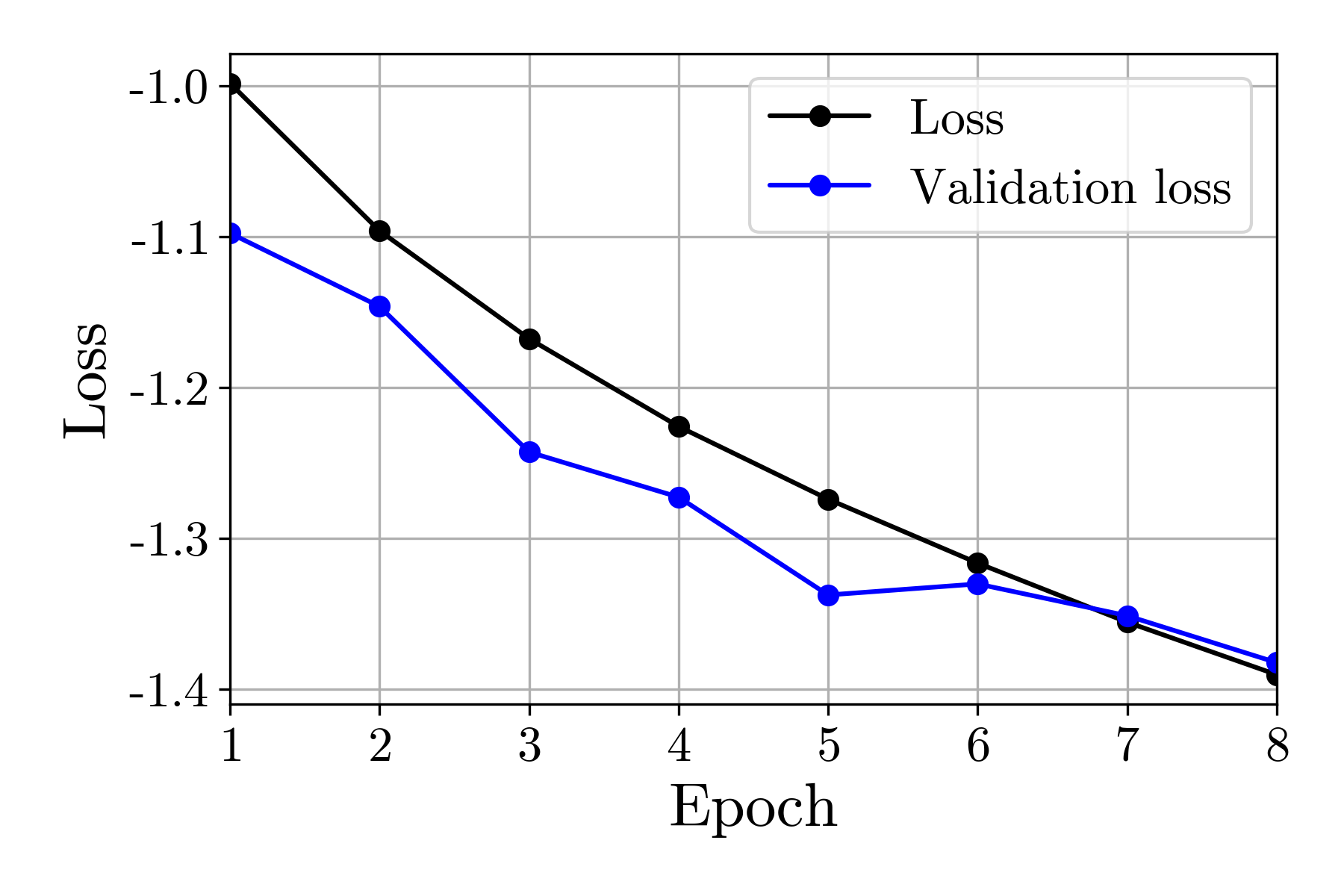}
    \caption{Metrics tracked during the training procedure.}
    \label{fig:TrainingMetrics}
\end{figure}

The evolution of the metrics tracked is displayed in Fig.~\ref{fig:TrainingMetrics}. The loss, as expected, decreases steadily, indicating that the CNN is continuously learning to fit the data. This indicates a good behavior over the training set. The validation loss also displays a decrease during the training indicating that the CNN is also learning how to generalize the results over data that has not seen before.

The final loss and validation loss that are obtained for the CNN chosen, the one of the eighth epoch, are both $-1.13$. This CNN is tested with the other data set and two real events in the following section.

It is worth noting that the CNN performance could be further improved by expanding the training set to include a wider range of signals and noise conditions. Moreover, a higher fine-tuning of the model's hyperparameters, such as adjusting the learning rate or the architecture, could potentially enhance its accuracy and reduce a bit more the uncertainties in the estimated parameters. This is out of the scope of the paper, as it intends to be a proof-of-concept.

\section{Results}
\label{sec:results}

\begin{figure}
	\includegraphics[width=\columnwidth]{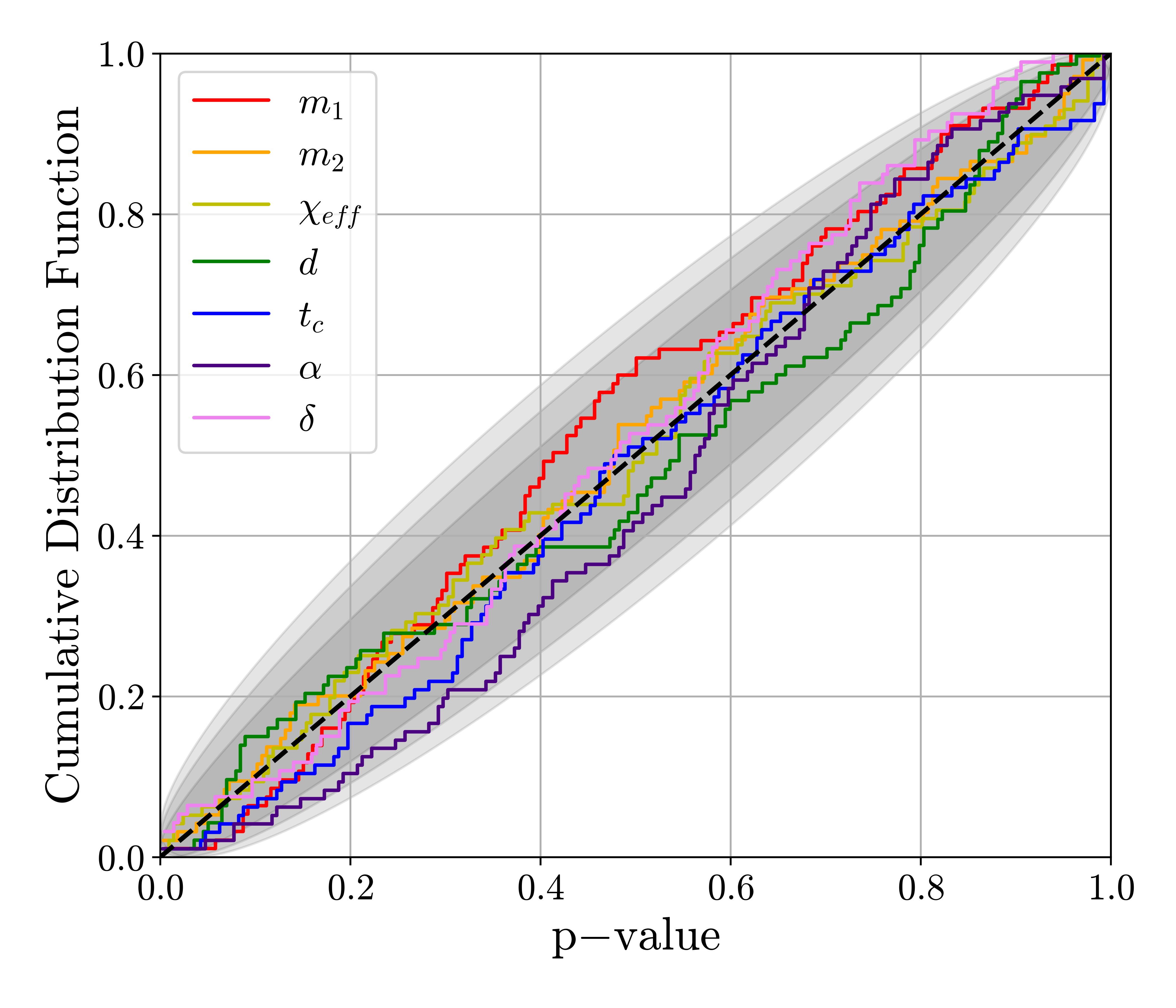}
    \caption{P-P plot for the different variables that are being fitted by the CNN. The individual masses are obtained by post-processing the chirp mass and mass ratio. The shaded regions indicate the $\pm 1\sigma$, $\pm 2\sigma$ and $\pm 3\sigma$ confidence levels in decreasing intensity of color, respectively.}
    \label{fig:ppPlot}
\end{figure}

In this final section, the main results of the CNN are presented. The first test that can be performed to ensure the proper performance is constructing the probability-probability (or P-P for short) plot. This plot is constructed by performing the inference of the CNN over the test set and evaluating the p-value of the true parameter. This is computed as
\begin{equation}
    \mathrm{p-value} = \int_{\theta_{\mathrm{true}}}^\infty f(\theta; \alpha, \beta)\td\theta~.
\end{equation}
 If the distribution outputted by the CNN is a true statistical distribution, the cumulative probability density of this p-value should follow a $45^\circ$ line between the points $(0,0)$ and $(1,1)$. This plot is shown in Fig.~\ref{fig:ppPlot}.

These cross-validation results demonstrate the robustness of our CNN approach, as it consistently produces a reliable parameter estimation using a new subset of data.

Since the CNN has been trained on simulated signals and on Gaussian noise it is interesting to test it using real data. The GWTC-3 catalog \citep{Abbott2021GWTC-3:Run} contains the $35$ O3b confident detections. From this collection, a judicious selection of events is needed following certain criteria. Namely, the chosen events must have been detected during periods when all three interferometers were actively online and when the data quality recorded across these interferometers was designated as optimal. Furthermore, compatibility with the CNN's training regimen requires that the parameters derived through conventional methodologies lie within the predetermined range calibrated for the neural network. Finally, only events exhibiting a SNR exceeding 10 are considered viable candidates as this was a decision during training. Accordingly, among these stringent criteria, the events GW200129\_065458, GW200224\_222234 and GW200311\_115853 emerge as good choices for this validation, fulfilling the requirements and having different parameters. 



\begin{figure*}
	\includegraphics[width=\textwidth]{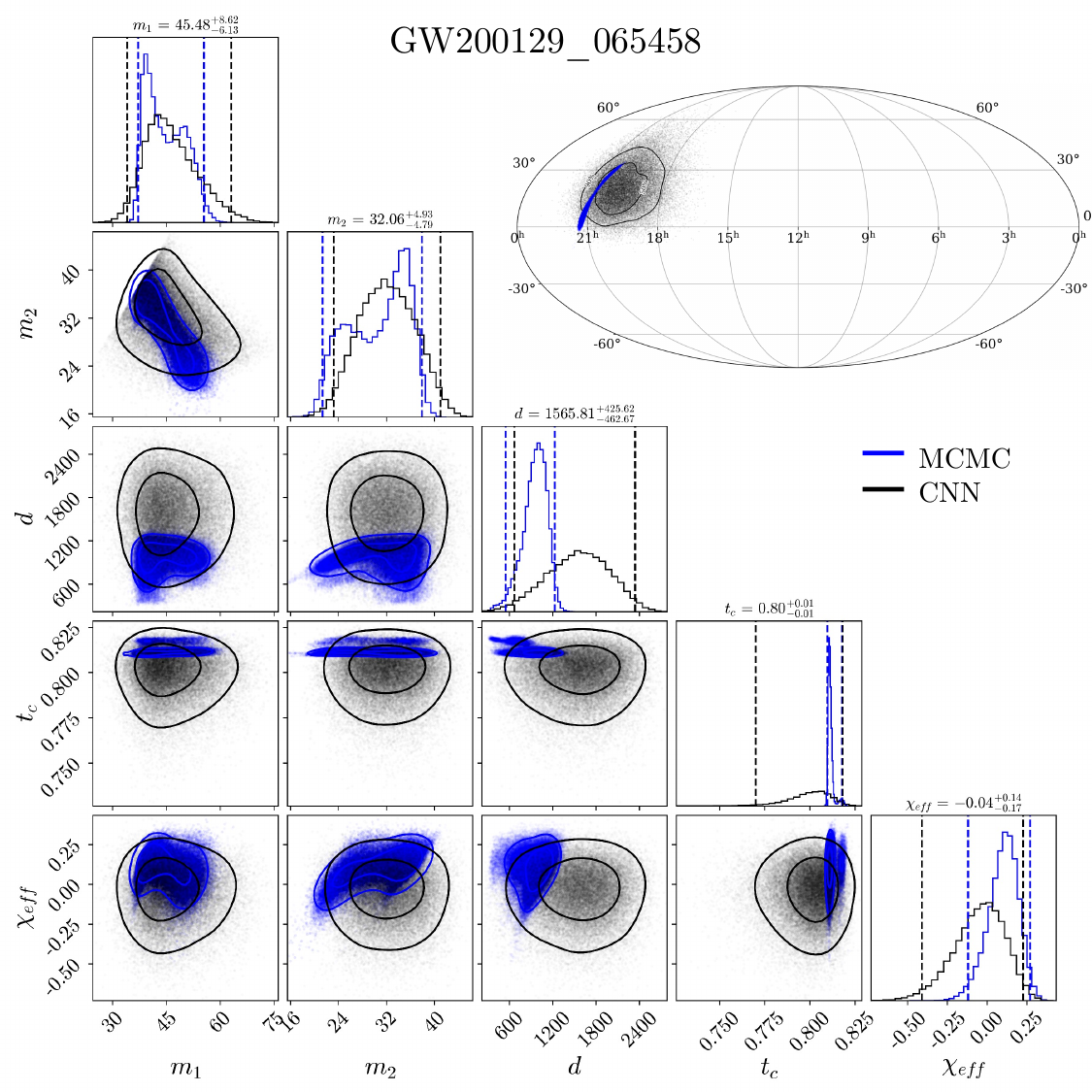}
    \caption{Full posterior distribution obtained for the CNN and for the public parameter estimation release obtained with a MCMC approach for the GW200129\_065458 event.}
    \label{fig:FullPosterior_GW200129_065458}
\end{figure*}

For GW200129\_065458, we apply the CNN on the data that contains it and compare it to the posterior distributions published in the GWTC-3 catalog by the LVK collaboration using the traditional Bayesian inference methods\citep{Abbott2021GWTC-3:Run,GWTC-3_OpenData}. The result is shown in Fig.~\ref{fig:FullPosterior_GW200129_065458}. To generate $10,000$ samples of the posterior the CNN, running on a GPU NVIDIA GeForce RTX 2080 Ti, takes $0.05$ s. This implies an improvement of several orders of magnitude with respect to the traditional method.

\begin{figure*}
	\includegraphics[width=\textwidth]{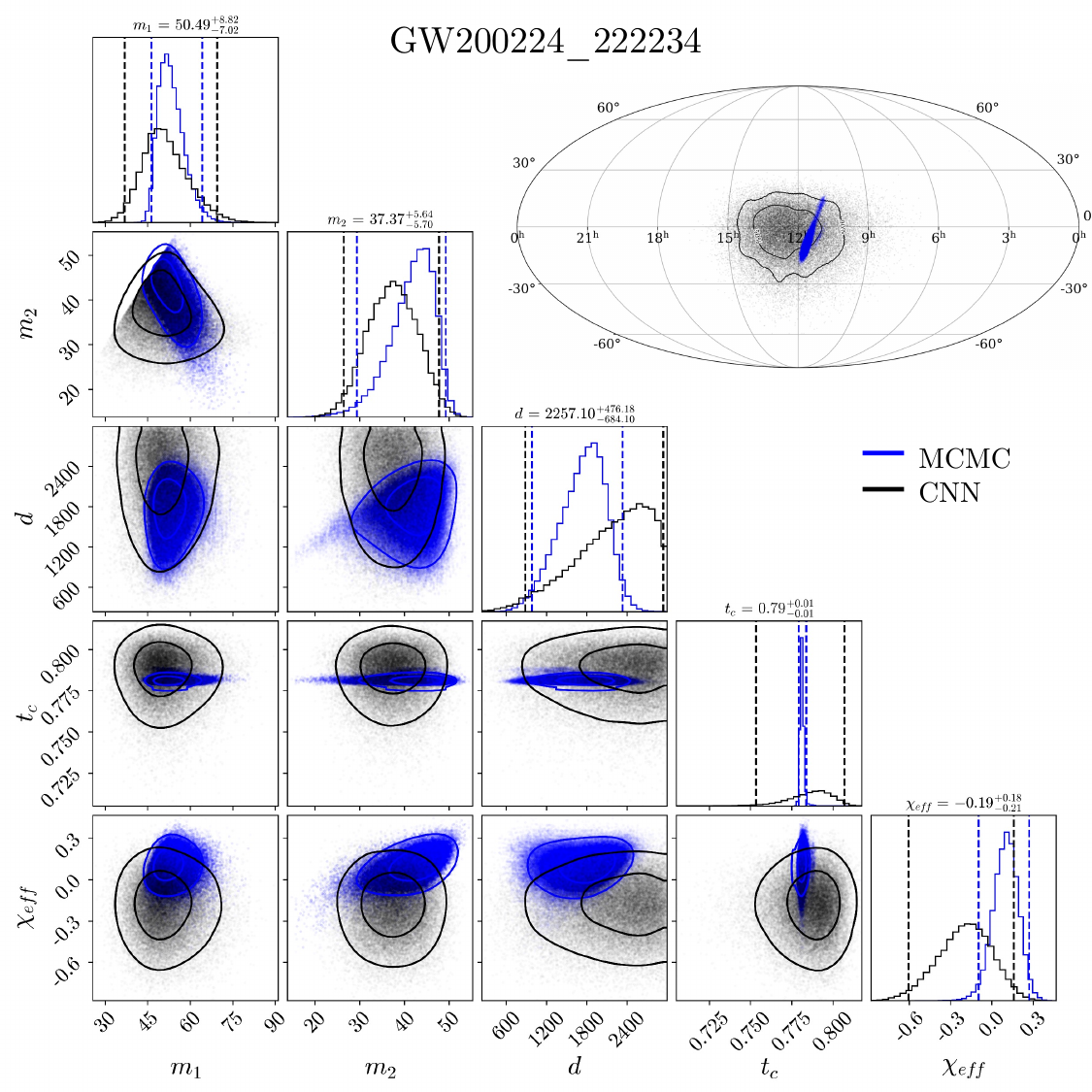}
    \caption{Full posterior distribution obtained for the CNN and for the public parameter estimation release obtained with a MCMC approach for the GW200224\_222234 event.}
    \label{fig:FullPosterior_GW200224_222234}
\end{figure*}

\begin{figure*}
	\includegraphics[width=\textwidth]{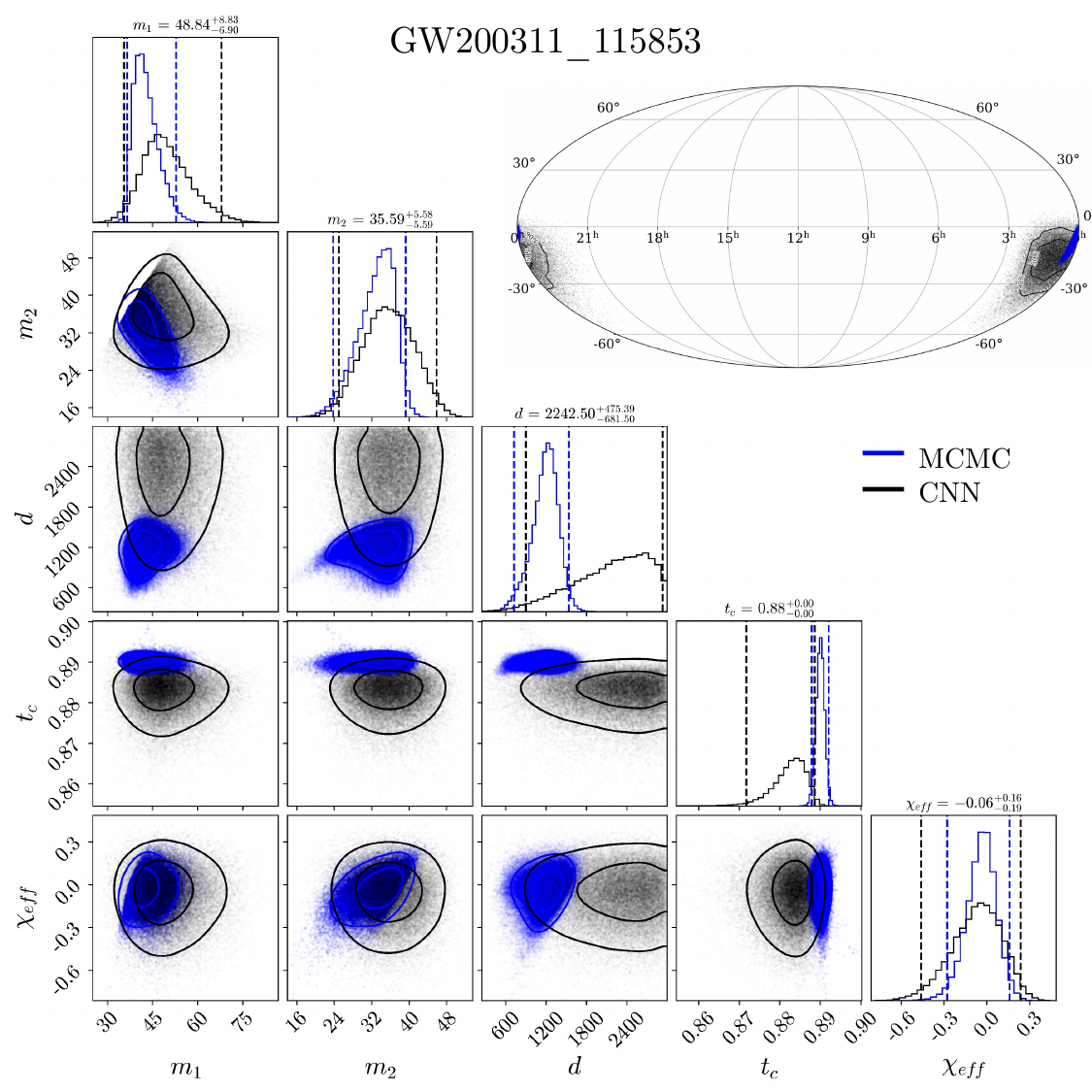}
    \caption{Full posterior distribution obtained for the CNN and for the public parameter estimation release obtained with a MCMC approach for the GW200311\_115853 event.}
    \label{fig:FullPosterior_GW200311_115853}
\end{figure*}

Our CNN is able to produce posterior distributions that in all cases are compatible with the published results. The one that exhibits the worst similarity is, for this particular event, the distance, as it overestimates it.  In this case, the sky position has a big overlap with the MCMC estimated one, but still has a larger uncertainty.

The results for the event GW200224\_222234 are shown in Fig.~\ref{fig:FullPosterior_GW200224_222234}. In this case, all the estimated parameters are also in accordance with those published in the GWTC-3 catalog. The one that exhibits the worst performance is, in this case, the effective spin.  Regarding the sky position, the uncertainty yielded by the CNN is too large to accurately pinpoint the event, but it still is unbiased and could produce a first alert for the instruments that look for electromagnetic counterparts to start pointing their instruments to a given patch in the sky while the more accurate pipelines improve the precision of the sky localization. 

A typical short gamma ray burst (GRB), as the one observed alongside GW170817 \citep{GW170817}, lasts less than $2$ s and, taking into account that the instruments might need time to point towards a given direction, ideally the inference speed should be of a fraction of a second to avoid becoming the bottleneck. This is achieved by our neural network.

Finally, the results for the GW200311\_115853 event are shown in Fig.~\ref{fig:FullPosterior_GW200311_115853}. In this particular example, the coalesce time is slightly underestimated and the distance overestimated. The rest of the parameters are well predicted but with larger uncertainties.

Overall, the results indicate a good response from CNN. An important advantage of our approach is its computational efficiency. The CNN model provides several posterior samples in a fraction of the time required by Bayesian inference methods. This enhanced capability enables real-time analysis of GW signals and even facilitates prompt alerts for possible follow-up observations and multi-messenger astronomy collaborations.

Despite the high inference speed, the CNN approach introduces a trade-off in terms of higher parameter uncertainty. These can be mainly attributed to the simplified assumptions made by the CNN architecture and the limited training data used. However, even with these uncertainties, the CNN model can serve as an initial indication for the sky position and other key parameters, enabling a swift response and triggering more accurate analyses by other slower, yet more accurate, pipelines.

In summary, with these results, our CNN model demonstrates promising capabilities for a fast parameter estimation of GWs. It provides a reliable posterior distribution, while exhibits a competitive performance compared to traditional Bayesian sampling methods, and offers real-time inference capabilities. This is only a proof-of-concept and future research can focus on refining the CNN architecture, incorporating additional data sources, and exploring ensembling techniques to further improve the accuracy and robustness of gravitational wave parameter estimation.

\section{Conclusions}

We have presented a neural network to perform the typically computationally expensive task of estimating the parameters of gravitational wave events. The chosen architecture has been a one-dimensional convolutional neural network with three channels, one per available interferometer, and predicting the parameters of a Kumaraswamy distribution. These parameters then define what can be regarded as the posterior distribution. The training shows that this architecture can correctly learn how to estimate them and that the outputted distribution behaves like a probability density function. Finally, it has been applied to real events, and the results have shown that, although our CNN has a larger uncertainty than traditional Bayesian inference approaches, the actual parameters are generally in agreement with the outputted distribution. Our approach offers real-time inference speeds that could be used for possible multi-messenger follow-ups. 

\section*{Acknowledgements}

 This project has received funding from the European Union’s Horizon 2020 research and innovation programme under the Marie Skłodowska-Curie Grant Agreement No. 754510. This work is partially supported by the Spanish MCIN/AEI/10.13039/501100011033 under the Grants
No. SEV-2016-0588, No. PGC2018-101858-B-I00, and No. PID2020-113701GB-I00, some
of which include ERDF funds from the European Union, and by the MICINN with funding from the European Union NextGenerationEU (PRTR-C17.I1) and by the Generalitat de
Catalunya. IFAE is partially funded by the CERCA program of the Generalitat de Catalunya.
MAC is supported by the 2022 FI-00335 grant. The corner plots use the \textit{corner.py} library \citep{corner}. This document has received a LIGO DCC number of P2300296 and Virgo TDS number of VIR-0791A-23. This research has made use of data, software, and/or web tools obtained from the Gravitational Wave Open Science Center (\url{https://www.gw-openscience.org/}), a service of the LIGO Laboratory, the LIGO Scientific Collaboration, and the Virgo Collaboration. LIGO Laboratory and Advanced LIGO are funded by the United States National Science Foundation (NSF) as well as the Science and Technology Facilities Council (STFC) of the United Kingdom, the Max-Planck-Society (MPS), and the State of Niedersachsen/Germany for support of the construction of Advanced LIGO and construction and operation of the GEO600 detector. Additional support for Advanced LIGO was provided by the Australian Research Council. Virgo is funded, through the European Gravitational Observatory (EGO), by the French Centre National de Recherche Scientifique (CNRS), the Italian Istituto Nazionale di Fisica Nucleare (INFN), and the Dutch Nikhef, with contributions by institutions from Belgium, Germany, Greece, Hungary, Ireland, Japan, Monaco, Poland, Portugal, and Spain.

\section*{Data Availability}

The codes and data generated for this paper are available under reasonable request to the atuhors.



\bibliographystyle{mnras}
\bibliography{references} 

\begin{thebibliography}{}
\makeatletter
\relax
\def\mn@urlcharsother{\let\do\@makeother \do\$\do\&\do\#\do\^\do\_\do\%\do\~}
\def\mn@doi{\begingroup\mn@urlcharsother \@ifnextchar [ {\mn@doi@}
  {\mn@doi@[]}}
\def\mn@doi@[#1]#2{\def\@tempa{#1}\ifx\@tempa\@empty \href
  {http://dx.doi.org/#2} {doi:#2}\else \href {http://dx.doi.org/#2} {#1}\fi
  \endgroup}
\def\mn@eprint#1#2{\mn@eprint@#1:#2::\@nil}
\def\mn@eprint@arXiv#1{\href {http://arxiv.org/abs/#1} {{\tt arXiv:#1}}}
\def\mn@eprint@dblp#1{\href {http://dblp.uni-trier.de/rec/bibtex/#1.xml}
  {dblp:#1}}
\def\mn@eprint@#1:#2:#3:#4\@nil{\def\@tempa {#1}\def\@tempb {#2}\def\@tempc
  {#3}\ifx \@tempc \@empty \let \@tempc \@tempb \let \@tempb \@tempa \fi \ifx
  \@tempb \@empty \def\@tempb {arXiv}\fi \@ifundefined
  {mn@eprint@\@tempb}{\@tempb:\@tempc}{\expandafter \expandafter \csname
  mn@eprint@\@tempb\endcsname \expandafter{\@tempc}}}

\bibitem[\protect\citeauthoryear{Aasi et~al.}{Aasi et~al.}{2015}]{AdvLIGO}
Aasi J.,  et~al., 2015, \mn@doi [Class. Quant. Grav.]
  {10.1088/0264-9381/32/7/074001}, 32, 074001

\bibitem[\protect\citeauthoryear{Abadi et~al.}{Abadi
  et~al.}{2016}]{abadi2016tensorflow}
Abadi M.,  et~al., 2016, in 12th USENIX symposium on operating systems design
  and implementation (OSDI 16). p.~265

\bibitem[\protect\citeauthoryear{Abbott et~al.}{Abbott
  et~al.}{2016}]{FirstGWDet}
Abbott B.,  et~al., 2016, \mn@doi [Phys. Rev. Lett.]
  {https://doi.org/10.1103/PhysRevLett.116.061102}, 116, 061102

\bibitem[\protect\citeauthoryear{Abbott et~al.}{Abbott et~al.}{2017}]{GW170817}
Abbott B.~P.,  et~al., 2017, \mn@doi [Phys. Rev. Lett.]
  {10.1103/PhysRevLett.119.161101}, 119, 161101

\bibitem[\protect\citeauthoryear{Abbott et~al.}{Abbott
  et~al.}{2019}]{Abbott2019GWTC-1:Runs}
Abbott R.,  et~al., 2019, \mn@doi [Phys. Rev. X]
  {https://doi.org/10.1103/PhysRevX.9.031040}, 9, 031040

\bibitem[\protect\citeauthoryear{Abbott et~al.}{Abbott
  et~al.}{2020}]{Abbott2020GWTC-2:Run}
Abbott R.,  et~al., 2020, \mn@doi [Phys. Rev. X]
  {https://doi.org/10.1103/PhysRevX.11.021053}, 11, 021053

\bibitem[\protect\citeauthoryear{Abbott et~al.}{Abbott
  et~al.}{2021a}]{Abbott2021GWTC-3:Run}
Abbott R.,  et~al., 2021a, \mn@doi [arXiv:2111.03606]
  {https://doi.org/10.48550/arXiv.2111.03606}

\bibitem[\protect\citeauthoryear{Abbott et~al.}{Abbott et~al.}{2021b}]{TGR}
Abbott R.,  et~al., 2021b, \mn@doi [arXiv:2112.06861]
  {https://doi.org/10.48550/arXiv.2112.06861}

\bibitem[\protect\citeauthoryear{Abbott et~al.}{Abbott et~al.}{2022a}]{CW3}
Abbott R.,  et~al., 2022a, \mn@doi [Phys. Rev. D]
  {10.1103/PhysRevD.106.042003}, 106, 042003

\bibitem[\protect\citeauthoryear{Abbott et~al.}{Abbott et~al.}{2022b}]{CW1}
Abbott R.,  et~al., 2022b, \mn@doi [Phys. Rev. D]
  {10.1103/PhysRevD.106.102008}, 106, 102008

\bibitem[\protect\citeauthoryear{Abbott et~al.}{Abbott et~al.}{2022c}]{CW2}
Abbott R.,  et~al., 2022c, \mn@doi [ApJ.] {10.3847/1538-4357/ac6acf}, 935, 1

\bibitem[\protect\citeauthoryear{Abbott et~al.}{Abbott
  et~al.}{2023a}]{GWTC-3_OpenData}
Abbott R.,  et~al., 2023a, \mn@doi [arXiv:2302.03676]
  {https://doi.org/10.48550/arXiv.2302.03676}

\bibitem[\protect\citeauthoryear{Abbott et~al.}{Abbott et~al.}{2023b}]{POP}
Abbott R.,  et~al., 2023b, \mn@doi [Phys. Rev. X] {10.1103/PhysRevX.13.011048},
  13, 011048

\bibitem[\protect\citeauthoryear{Acernese et~al.}{Acernese
  et~al.}{2015}]{AdvVIRGO}
Acernese F.,  et~al., 2015, \mn@doi [Class. Quantum Grav.]
  {10.1088/0264-9381/32/2/024001}, 32, 024001

\bibitem[\protect\citeauthoryear{Alvey et~al.}{Alvey
  et~al.}{2023}]{Alvey:2023naa}
Alvey J.,  et~al., 2023, \mn@doi [arXiv:2308.06318] {10.48550/arXiv.2308.06318}

\bibitem[\protect\citeauthoryear{Andr\'es-Carcasona et~al.}{Andr\'es-Carcasona
  et~al.}{2023}]{Andres-Carcasona:2022}
Andr\'es-Carcasona M.,  et~al., 2023, \mn@doi [Phys. Rev. D]
  {10.1103/PhysRevD.107.082003}, 107, 082003

\bibitem[\protect\citeauthoryear{Ashton et~al.}{Ashton et~al.}{2019}]{Bilby}
Ashton G.,  et~al., 2019, \mn@doi [ApJS] {10.3847/1538-4365/ab06fc}, 241, 27

\bibitem[\protect\citeauthoryear{Berry et~al.}{Berry et~al.}{2015}]{Berry:2014}
Berry C. P.~L.,  et~al., 2015, \mn@doi [ApJ] {10.1088/0004-637X/804/2/114},
  804, 114

\bibitem[\protect\citeauthoryear{Bhardwaj et~al.}{Bhardwaj
  et~al.}{2023}]{Bhardwaj2023}
Bhardwaj U.,  et~al., 2023, \mn@doi [Phys. Rev. D]
  {10.1103/PhysRevD.108.042004}, 108, 042004

\bibitem[\protect\citeauthoryear{Chua \& Vallisneri}{Chua \&
  Vallisneri}{2020}]{Chua:2019}
Chua A. J.~K.,  Vallisneri M.,  2020, \mn@doi [Phys. Rev. Lett.]
  {10.1103/PhysRevLett.124.041102}, 124, 041102

\bibitem[\protect\citeauthoryear{Crisostomi et~al.}{Crisostomi
  et~al.}{2023}]{Crisostomi_2023}
Crisostomi M.,  et~al., 2023, \mn@doi [Phys. Rev. D]
  {10.1103/PhysRevD.108.044029}, 108, 044029

\bibitem[\protect\citeauthoryear{Cuoco et~al.}{Cuoco
  et~al.}{2020}]{cuoco2020enhancing}
Cuoco E.,  et~al., 2020, \mn@doi [Mach. Learn.: Sci. Technol.]
  {https://doi.org/10.1088/2632-2153/abb93a}, 2, 011002

\bibitem[\protect\citeauthoryear{Dax et~al.}{Dax et~al.}{2023}]{DaxDingo}
Dax M.,  et~al., 2023, \mn@doi [Phys. Rev. Lett.]
  {10.1103/PhysRevLett.130.171403}, 130, 171403

\bibitem[\protect\citeauthoryear{{Dillon} et~al.}{{Dillon}
  et~al.}{2017}]{TensorflowProbability}
{Dillon} J.~V.,  et~al., 2017, \mn@doi [arXiv:1711.10604]
  {https://doi.org/10.48550/arXiv.1711.10604}

\bibitem[\protect\citeauthoryear{Fan et~al.}{Fan et~al.}{2019}]{Fan:2018}
Fan X.,  et~al., 2019, \mn@doi [Sci. China Phys. Mech. Astron.]
  {10.1007/s11433-018-9321-7}, 62, 969512

\bibitem[\protect\citeauthoryear{Foreman-Mackey}{Foreman-Mackey}{2016}]{corner}
Foreman-Mackey D.,  2016, \mn@doi [JOSS] {10.21105/joss.00024}, 1, 24

\bibitem[\protect\citeauthoryear{Gabbard et~al.}{Gabbard
  et~al.}{2018}]{Gabbard:2017}
Gabbard H.,  et~al., 2018, \mn@doi [Phys. Rev. Lett.]
  {10.1103/PhysRevLett.120.141103}, 120, 141103

\bibitem[\protect\citeauthoryear{Gabbard et~al.}{Gabbard
  et~al.}{2022}]{Gabbard:2019}
Gabbard H.,  et~al., 2022, \mn@doi [Nat. Phys.] {10.1038/s41567-021-01425-7},
  18, 112

\bibitem[\protect\citeauthoryear{George \& Huerta}{George \&
  Huerta}{2018}]{George:2017_DL}
George D.,  Huerta E.~A.,  2018, \mn@doi [Phys. Lett. B]
  {10.1016/j.physletb.2017.12.053}, 778, 64

\bibitem[\protect\citeauthoryear{George et~al.}{George
  et~al.}{2018}]{George:2018}
George D.,  et~al., 2018, \mn@doi [Phys. Rev. D] {10.1103/PhysRevD.97.101501},
  97, 101501

\bibitem[\protect\citeauthoryear{Gilks}{Gilks}{2005}]{Gilks_MCMC}
Gilks W.~R.,  2005, Markov Chain Monte Carlo.
John Wiley \& Sons, Ltd, \mn@doi{https://doi.org/10.1002/0470011815.b2a14021}

\bibitem[\protect\citeauthoryear{Green \& Gair}{Green \&
  Gair}{2021}]{Green:2020GW150914}
Green S.~R.,  Gair J.,  2021, \mn@doi [Mach. Learn.: Sci. Technol.]
  {10.1088/2632-2153/abfaed}, 2, 03LT01

\bibitem[\protect\citeauthoryear{Green et~al.}{Green et~al.}{2020}]{Green:2020}
Green S.~R.,  et~al., 2020, \mn@doi [Phys. Rev. D]
  {10.1103/PhysRevD.102.104057}, 102, 104057

\bibitem[\protect\citeauthoryear{Husa et~al.}{Husa
  et~al.}{2016}]{IMRPhenomPv2_1}
Husa S.,  et~al., 2016, \mn@doi [Phys. Rev. D] {10.1103/PhysRevD.93.044006},
  93, 044006

\bibitem[\protect\citeauthoryear{Khan et~al.}{Khan
  et~al.}{2016}]{IMRPhenomPv2_2}
Khan S.,  et~al., 2016, \mn@doi [Phys. Rev. D] {10.1103/PhysRevD.93.044007},
  93, 044007

\bibitem[\protect\citeauthoryear{Krastev et~al.}{Krastev
  et~al.}{2021}]{Krastev:2020}
Krastev P.~G.,  et~al., 2021, \mn@doi [Phys. Lett. B]
  {10.1016/j.physletb.2021.136161}, 815, 136161

\bibitem[\protect\citeauthoryear{Kumaraswamy}{Kumaraswamy}{1980}]{Kumaraswamy}
Kumaraswamy P.,  1980, \mn@doi [J. Hydrol.]
  {https://doi.org/10.1016/0022-1694(80)90036-0}, 46, 79

\bibitem[\protect\citeauthoryear{Men{\'{e}}ndez-V{\'{a}}zquez
  et~al.}{Men{\'{e}}ndez-V{\'{a}}zquez
  et~al.}{2021}]{Menendez-Vazquez2021SearchesPeriod}
Men{\'{e}}ndez-V{\'{a}}zquez A.,  et~al., 2021, \mn@doi [Phys. Rev. D]
  {10.1103/PhysRevD.103.062004}, 103, 062004

\bibitem[\protect\citeauthoryear{Morr\'as et~al.}{Morr\'as
  et~al.}{2022}]{Morras:2021}
Morr\'as G.,  et~al., 2022, \mn@doi [Phys. Dark Universe]
  {10.1016/j.dark.2021.100932}, 35, 100932

\bibitem[\protect\citeauthoryear{Nitz et~al.}{Nitz
  et~al.}{2020}]{nitz2020gwastro}
Nitz A.,  et~al., (2020), gwastro/pycbc: PyCBC release v1. 16.11,
  \url{https://zenodo.org/record/4134752/export/json#.ZBr_7ezMJqs}

\bibitem[\protect\citeauthoryear{Pankow et~al.}{Pankow
  et~al.}{2015}]{Pankow:2015}
Pankow C.,  et~al., 2015, \mn@doi [Phys. Rev. D] {10.1103/PhysRevD.92.023002},
  92, 023002

\bibitem[\protect\citeauthoryear{Romero-Shaw et~al.}{Romero-Shaw
  et~al.}{2020}]{Bilby2}
Romero-Shaw I.~M.,  et~al., 2020, \mn@doi [MNRS] {10.1093/mnras/staa2850}, 499,
  3295

\bibitem[\protect\citeauthoryear{Singer \& Price}{Singer \&
  Price}{2016}]{Singer:2015_RapidBayesian}
Singer L.~P.,  Price L.~R.,  2016, \mn@doi [Phys. Rev. D]
  {10.1103/PhysRevD.93.024013}, 93, 024013

\bibitem[\protect\citeauthoryear{Skilling}{Skilling}{2006}]{NestedSampling}
Skilling J.,  2006, \mn@doi [Bayesian Anal.] {10.1214/06-BA127}, 1, 833

\bibitem[\protect\citeauthoryear{Smith et~al.}{Smith
  et~al.}{2020}]{Smith:2019rapidBayesian}
Smith R. J.~E.,  et~al., 2020, \mn@doi [MNRAS] {10.1093/mnras/staa2483}, 498,
  4492

\bibitem[\protect\citeauthoryear{Usman et~al.}{Usman
  et~al.}{2016}]{Usman2016TheCoalescence}
Usman S.~A.,  et~al., 2016, \mn@doi [Class. Quantum Grav.]
  {10.1088/0264-9381/33/21/215004}, 33, 215004

\bibitem[\protect\citeauthoryear{Veitch et~al.}{Veitch
  et~al.}{2015}]{Veitch_LALinference}
Veitch J.,  et~al., 2015, \mn@doi [Phys. Rev. D] {10.1103/PhysRevD.91.042003},
  91, 042003

\bibitem[\protect\citeauthoryear{Zevin et~al.}{Zevin et~al.}{2017}]{GWspy}
Zevin M.,  et~al., 2017, \mn@doi [Class. Quantum Grav.]
  {10.1088/1361-6382/aa5cea}, 34, 064003

\makeatother
\end{thebibliography}








\bsp	
\label{lastpage}
\end{document}